\documentclass[pdflatex,mathphys]{sty}

\usepackage{upgreek}
\usepackage{siunitx}
\usepackage{gensymb}
\jyear{2025}%

\unnumbered

\begin{document}

\title[Photonic neuron by Modulation-and-weighting microring]{Compact, Reconfigurable, and Scalable Photonic Neurons by Modulation-and-Weighting Microring Resonators}

\author*[1,2]{\fnm{Weipeng} \sur{Zhang}}\email{weipeng@lightxcelerate.com}
\equalcont{These authors contributed equally to this work.}
\author[1]{\fnm{Yuxin} \sur{Wang}}\email{yuxinw@princeton.edu}
\equalcont{These authors contributed equally to this work.}
\author[1]{\fnm{Joshua C.} \sur{Lederman}}\email{joshuacl@princeton.edu}
\author[3]{\fnm{Bhavin J.} \sur{Shastri}}\email{shastri@ieee.org}
\author*[1]{\fnm{Paul R.} \sur{Prucnal}}\email{prucnal@princeton.edu}

\affil*[1]{\orgdiv{Department of Electrical and Computer Engineering}, \orgname{Princeton University}, \orgaddress{\city{Princeton}, \postcode{08544}, \state{New Jersey}, \country{USA}}}

\affil[2]{\orgname{LightXcelerate, Inc.}, \orgaddress{\city{Palo Alto}, \postcode{94301}, \state{California}, \country{USA}}}

\affil[3]{\orgdiv{Department of Physics, Engineering Physics and Astronomy}, \orgname{Queen’s University}, \orgaddress{\city{Kingston}, \postcode{K7L 3N6}, \state{Ontario}, \country{Canada}}}

\maketitle
\clearpage

\section{Abstract}
Neuromorphic photonics promises sub-nanosecond latency, ultrawide bandwidth, and high parallelism, but practical scalability is constrained by fabrication tolerances, spectral alignment, and tuning energy. Here, we present a large-scale, compact, and reconfigurable photonic neuron in which each microring performs modulation and weighting simultaneously. By exploiting both carrier and thermal tuning within a single device, this architecture reduces footprint, relaxes spectral alignment requirements to just two optical components, and yields a steep transfer response that lowers tuning energy. The proposed neuron supports multiple operating configurations, allowing its dynamical behavior to be adapted to different computational tasks. In particular, a short electrical feedback path enables recurrent operation, providing tunable short- and long-term memory for temporal processing. Using a 10-microring resonator array, we demonstrate both spatial and temporal computing, including a 3$\times$3 convolution for image processing with an error of $<$5\% and high-frequency financial time-series prediction. Each modulation-weighting element occupies 80$\times$45 \SI{}{\micro\meter^2} and consumes an average of \SI{0.186}{\milli\watt}, corresponding to a compute density of \SI{4.67}{TOPS/s/\milli\meter^2}. Excluding electronic power, the on-chip tuning efficiency reaches approximately \SI{105}{TOPs/\watt}, which is comparable to state-of-the-art implementations. These results indicate that modulation-and-weighting microring resonator banks provide a scalable building block for large-scale neuromorphic photonic systems, offering a favorable combination of compact footprint, low power consumption, and functional flexibility.

\section{Keywords}
Silicon Photonics, Neuromorphic Photonics, Photonic Neural Network

\section{1 Introduction}
Photonic computing offers sub-nanosecond latency, wide bandwidth, and intrinsic parallelism, making it a compelling platform for high-performance information processing. These attributes support accelerated AI training and real-time multimodal inference, with demonstrated relevance across applications ranging from genomics to climate modeling \cite{brunner2025roadmap, zhang2024system, shekhar2024roadmapping, huang2022prospects}. As a result, neuromorphic photonic architectures have attracted significant interest as candidates for next-generation intelligent computing. Even with this promise, their scalability has remained constrained \cite{fu2024optical, xu2023integrated, ferreira2017progress}. Photonic components cannot be arbitrarily miniaturized because their dimensions are fundamentally tied to the wavelength of light, limiting achievable device density. In addition, even slight process variances introduced during fabrication can cause substantial performance discrepancies, making it difficult to maintain uniformity across nominally identical devices \cite{xu2024control, ferreira2022design, bogaerts2018silicon, tait2017neuromorphic, bogaerts2013design}. Resonator-based components, such as microring resonators (MRRs), are susceptible to these imperfections despite their favorable footprint \cite{zhang2022silicon, tait2016microring, bogaerts2012silicon}. Collectively, these factors have impeded the deployment of neuromorphic photonic processors at the scale required by modern large AI models, which demand both complexity and reliability \cite{vaswani2017attention, kaplan2020scaling}.

A typical photonic processor requires spectral alignment among several key components, including the laser source, modulator, weighting element, and photodetector (PD) \cite{huang2020demonstration, tait2016microring}. While PDs generally operate over a broad wavelength range, aligning the remaining components simultaneously can be challenging, particularly when multiple MRRs exhibit narrow and device-specific resonant wavelengths. Using tunable lasers can mitigate alignment issues for one MRR element (either the modulator or the weighting element). However, alignment becomes significantly more difficult when both the modulator and weighting element rely on independently fabricated MRRs that are not spectrally matched. Such mismatches are common in practice due to fabrication tolerances and environmental drifts \cite{ferreira2022design, zhang2022silicon}. As the number of MRRs increases, the cumulative alignment burden grows rapidly, imposing a practical limit on the scalability of photonic processors.

\begin{figure}[ht!]
  \centering
  \includegraphics[width=.99\linewidth]{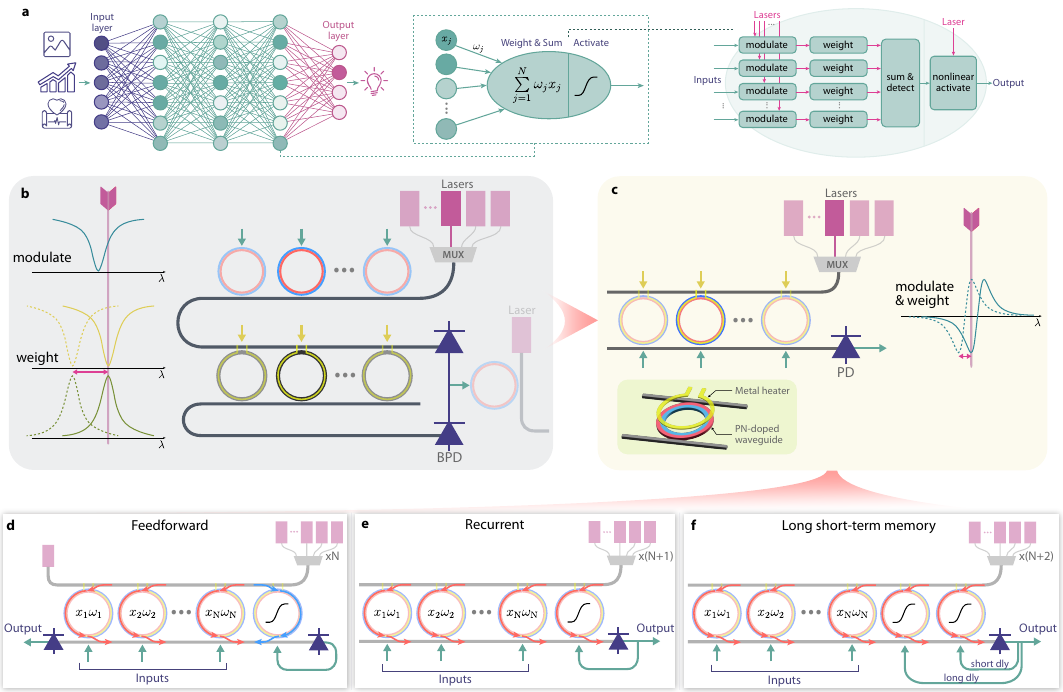}
\caption{(a) Neural network architecture composed of numerous small units that perform weighting, summation, and nonlinear activation. Photonic implementations typically require lasers and multiple integrated photonic devices for each function. (b) A typical MRR-based photonic neuron, comprising a bank of multiple lasers at different wavelengths, a ring modulator bank, an MRR-based weight bank, and a balanced PD. (c) The proposed approach, which consolidates both modulation and weighting in a single MRR bank and employs a single-ended PD, thereby reducing complexity and footprint. (d) Three possible configurations of the proposed MRR bank, enabling neuron types in feedforward, recurrent (short-term memory), and combined long- and short-term memory modes.
}
\label{fig:vision}
\end{figure}

To address these challenges, we propose a photonic neuron that integrates modulation and weighting within a single microring resonator. This unified design reduces the spectral alignment requirement from three active components to two, enabling proper alignment with a single tunable laser and avoiding the need to individually match multiple MRR-based elements. By leveraging both thermo-optic and carrier-based tuning within the same MRR, the device footprint is minimized. Notably, the resulting modulation-and-weighting MRR also exhibits a steeper transfer function between tuning current and effective weight than conventional weighting-only MRRs \cite{zhang2024microring}, leading to substantially improved tuning efficiency. The proposed MRR bank further supports recurrent operation through a simple electrical feedback path, providing tunable short- and long-term memory without introducing additional photonic components. The intrinsic nonlinear transfer characteristic of the MRR, whose shape and strength are tunable via the bias point, enables controllable nonlinearity in the recurrent configuration, a key requirement for effective neural network operation across diverse tasks. This reconfigurability allows the same hardware platform to support both spatial and temporal computing without chip redesign.

We demonstrate the scalability of the architecture using a 10-MRR array configured as a 3×3 convolutional kernel \cite{li2021survey, xu202111} for image processing tasks, including blurring, directional edge detection, and isotropic filtering. Beyond spatial filtering, we explore feedforward and recurrent configurations applied to real-world financial time-series data, illustrating the suitability of the platform for low-latency temporal analytics. Together, these results establish a hardware-level photonic neuron that combines architectural reconfigurability with high compute density and low power consumption, addressing key scalability constraints in neuromorphic photonic computing.

\section{2 Results}
\subsection{2.1 Modulation-and-weighting Microring Resonator}
We implemented two tuning mechanisms to enable both modulation and weighting within a single MRR. Specifically, there is PN doping in the ring waveguide, which leverages plasma dispersion effects (carrier depletion and injection), and an integrated metal heater on the ring, which provides thermo-optical tuning \cite{bogaerts2012silicon}. Both mechanisms modify the refractive index, thereby shifting the resonance and adjusting the transmitted optical power. Owing to its high bandwidth, the plasma dispersion effect is well-suited for modulating high-frequency input signals. In contrast, thermal tuning, though slower, offers a broader tuning range, making it ideal for setting the bias point.

Based on the small-signal model of the PN junction, the conversion efficiency is defined by the slope of the MRR transmission profile, which varies substantially near the resonance peak when these mechanisms operate in tandem. As a result, only a slight bias adjustment is required to shift the weight from its maximum positive setting to its maximum negative setting, enabling highly energy-efficient weight tuning. Moreover, integrating these two tuning approaches is fully compatible with standard silicon photonics foundry processes and is technically mature, requiring minimal fabrication process adjustments.

Mathematically, the thermal bias determines the effective synaptic weight by selecting the local slope of the Lorentzian transmission profile ($T$) of the MRR. In our modulation-and-weighting scheme, the weight corresponds to the small-signal modulation efficiency at the operating point and is therefore proportional to $dT/d\lambda\vert_\mathrm{bias}$. The resonance wavelength shifts with the heater according to $\lambda_0(i_{ht})=\lambda_0^{(0)}+\alpha_{th}i_{ht}$, which uniquely determines the detuning between the laser wavelength and the ring ($\alpha_{th}$ and $i_{ht}$ are the thermal tuning coefficient and tuning current). Linearizing the electro-optic modulation around this bias point yields the small-signal gain
\begin{equation}
    \omega(i_{ht})=K\left.\frac{dT}{d\lambda}\right\vert_{\lambda_L-\lambda_0(i_{ht})}
\end{equation}
where $\lambda_L$ denotes the laser wavelength and $K$ collects the optical input power and electrical-to-optical conversion factor. Supplementary Fig. S2 shows the measured weight–current characteristics, confirming that the weight is fully set by the Lorentzian slope selected by the thermal and carrier biases.

Because the MRR weights operate on the steep Lorentzian slope, modulation efficiency and tuning power are set by the available linear dynamic range. At our chosen operating points, the modulation-induced resonance shift remains in the small-signal regime, where the quadratic term contributes only a few percent distortion, which is equivalent to about 6 bits of adequate linearity. Furthermore, in-situ training automatically biases each MRR to adapt the intrinsic Lorentzian curvature, absorbing any residual nonlinearity. A detailed linearity analysis is provided in Supplementary Material. Weight stability is governed by detuning drift, which is suppressed by TEC-based temperature control and can be corrected by periodic recalibration or dithering-based stabilization techniques previously shown to achieve up to 9-bit precision \cite{zhang2022silicon}. The hysteresis is negligible under our operating conditions, as the optical power and PN drive remain below the bistability threshold, yielding single-valued and monotonic tuning curves as shown in Supplementary Fig. S3.

\begin{figure}[ht!]
  \centering
  \includegraphics[width=.99\linewidth]{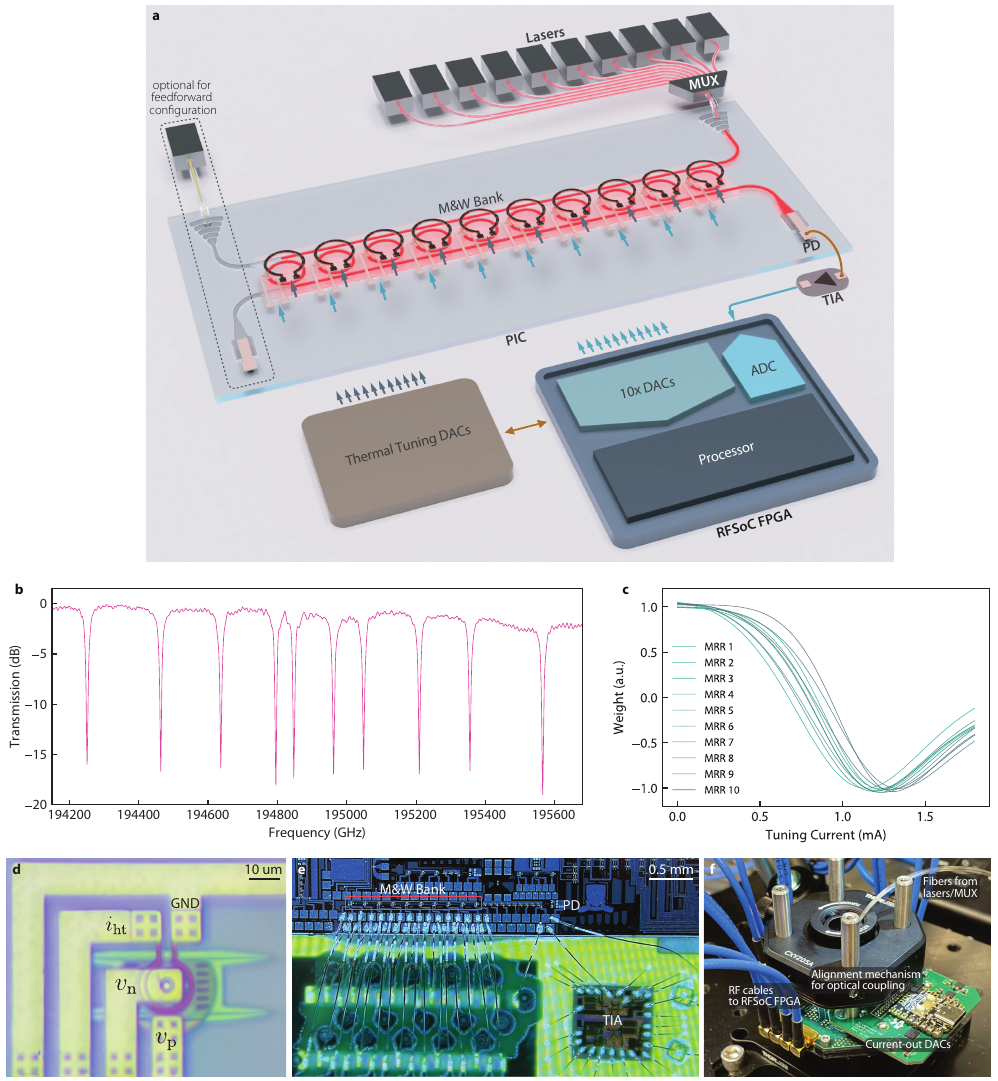}
\caption{Architecture and characterization of the large-scale photonic processor. (a) System schematic. (b) Optical spectra and (c) Current-tuning curve of the ten MRRs. (d) Micrograph of the dual-modulation microring. $i_{ht}$ denotes the thermal tuning current applied to the heater. $v_n$ and $v_p$ stand for the voltage applied to the p-doped and n-doped regions of the PN junction for modulation, respectively. (e) Micrograph of the photonic chip. (f) Photograph of the fully packaged chip.
}
\label{fig:setup}
\end{figure}

In addition to enabling both modulation and weighting, the microring resonator also provides the nonlinear activation essential for neuronal operation. The origin of this nonlinearity is inherent in the Lorentzian transmission profile of the microring. When the resonator operates in the small-signal limit, the input electrical modulation probes only a locally linear portion of the curve. However, as the modulation amplitude increases or as the DC bias detunes the operating point toward steeper regions of the resonance, the resulting electro-optic response becomes strongly nonlinear. Importantly, this nonlinear response is fully reconfigurable through bias tuning. By selecting different bias offsets relative to the resonance peak, the same device can therefore emulate sigmoid-like, ReLU-like, or quadratic activation behaviors, consistent with prior demonstrations of silicon-photonic neurons based on MRR modulators \cite{tait2019silicon}. The measured input–output transfer curves in Supplementary Fig. S1 confirm these behaviors and illustrate the range of nonlinear activation functions achievable through MRR bias tuning under typical operating conditions. Downstream of this nonlinearity, the PD operates strictly in its linear regime, converting the weighted optical sum into voltage without introducing any electronic or comparator-based nonlinearities. As a result, the entire multiply–sum–activate sequence is executed entirely in the analog domain with low latency.

As shown in Fig. \ref{fig:setup}(a), we fabricated a modulation-and-weighting bank comprising ten such MRRs. Each MRR has a slightly different radius, ranging from 8 to \SI{8.1}{\micro\meter}, resulting in a spectral spacing of approximately 51 GHz (equivalent to \SI{0.4}{\nano\meter}) between adjacent resonances (Fig. \ref{fig:setup}(b)). From this curve, we can infer that the linewidth is 3.2 GHz. Then, a Lorentzian-overlap estimate shows that inter-channel crosstalk is about 0.1\% (-30dB) in power, consistent with the absence of any observable interference in our experiments. Thermal cross-coupling was characterized by sweeping the heater of a single MRR while monitoring its nearest neighbor, resulting in a $<$5\% variation across the full weight range (see Supplementary Fig. S6).

The chip was fully packaged, as illustrated in Fig. \ref{fig:setup}(e) and (f), with a printed circuit board (PCB) as the interposer and wire-bonding of all the used pads. To drive the metal heater on each MRR, we integrated multi-channel current-output digital-to-analog converters, and impedance-controlled PCB traces to provide high-speed input signals to the PN junctions. The drop port of this modulation-and-weighting bank is connected to a PD, which is wire-bonded to a transimpedance amplifier for signal readout. The modulation bandwidth of the PN junction is tested to be 5.4 GHz, as shown in Supplementary Fig. S7.

The system is robust to fabrication variations and thermal drift through device tunability and system-level stabilization. The photonic integrated chip (PIC) in this work was fabricated using a 193 nm lithography process, which offers limited dimensional precision relative to the optical wavelength. This explains the observed difference between Fig. \ref{fig:setup}(b) and the simulated spectra in Supplementary Fig. S5. Although fabrication-induced resonance offsets exist, each channel uses an independently tunable laser that can be aligned to its measured resonance, eliminating the need for post-fabrication trimming. Thermal drift is minimized by a TEC, providing $\pm0.1\degree\mathrm{C}$ stability over hours. We also periodically refresh the current–weight mapping by re-sweeping each MRR to correct for slow drifts. With TEC control, per-channel laser tuning, and routine recalibration, the architecture remains tolerant to resonance shifts without requiring phase-change material or thermal-annealing trimming.

Regarding power consumption, Fig. \ref{fig:setup}(c) shows that all MRRs can tune throughout the range with only \SI{1.3}{\milli\ampere} of current, which translates to \SI{0.55}{\milli\watt} given that the metal heaters have \SI{330}{\ohm} resistance. Assuming a uniform distribution of the actual implemented weights in the setup, the average power across all MRRs is estimated at 0.186 mW. For comparison, a conventional MRR for weighting typically consumes \SI{10}{\milli\watt} for the embedded metal heater and \SI{3}{\milli\watt} for the doped waveguide heater \cite{tait2022quantifying}. Our architecture significantly reduces energy consumption compared to traditional designs with separated modulators and weighting elements, where thermal tuning can dominate the energy budget \cite{narayana2017morphonoc}. This reported power refers only to on-chip tuning, namely the thermo-optic bias of each MRR. This is the intrinsic power of the photonic weights and excludes system-level electronics such as FPGA logic, analog-to-digital converters (ADC), digital-to-analog converters (DAC), and transimpedance amplifiers (TIA). In addition to reduced power, the system achieves a fully integrated photonic processor with a complete signal-processing path, from input modulation to output detection, thereby minimizing processing latency. A full system-level power and latency breakdown is provided in Supplementary Material.

\subsection{2.2 Compact and Configurable Photonic Neuron}

The integration of modulation and weighting within a single MRR reduces the footprint, enabling more MRRs to be incorporated per neuron and thereby supporting a broader range of configurable architectures. The most common type, the feedforward neuron, performs weighted summation using only current inputs; after nonlinear activation, the output is propagated downstream. However, in many scenarios, performance improves by incorporating historical inputs. In our proposed photonic neuron, such a memory effect can be realized by re-routing the neuron output back to its input with a tunable delay. This approach is exceptionally straightforward in large-scale neurons, where a sufficient number of MRRs makes it easy to dedicate a few to processing historical data. With ten MRRs available, a selected subset can process real-time inputs, while the remainder handle recurrent (historical) data.

As illustrated in Fig. \ref{fig:vision}(d), we present two such recurrent implementations that use one or two MRRs for historical data. By adjusting the path length from the PD output to the MRR input, we can fine-tune the delay and thereby accommodate both long- and short-term memory effects. Specifically, we implemented the delay on the electrical signal path. A high-speed TIA amplifies the photocurrent from the on-chip detector, and the resulting voltage is routed back to the selected MRR input through a length-calibrated coaxial cable, achieving the delay that matches the corresponding data rate. This simple electrical loop provides a stable and sufficiently accurate delay for the recurrent prediction task. Further details are included in Supplementary Material.

This flexibility is especially advantageous for time-series data, such as medical information \cite{dong2023higher} and radio-frequency (RF) signals \cite{zhang2024system, lederman2023real, zhang2023broadband}, where crucial hidden features may only emerge when both current and previous inputs are considered. In the following sections, we first evaluate the proposed neuron in a convolutional image task that processes only recent inputs, and subsequently demonstrate the benefits of its memory capability in a financial time-series context.

\subsection{2.3 Demonstration of Image Processing}

\begin{figure}[ht!]
  \centering
  \includegraphics[width=.8\linewidth]{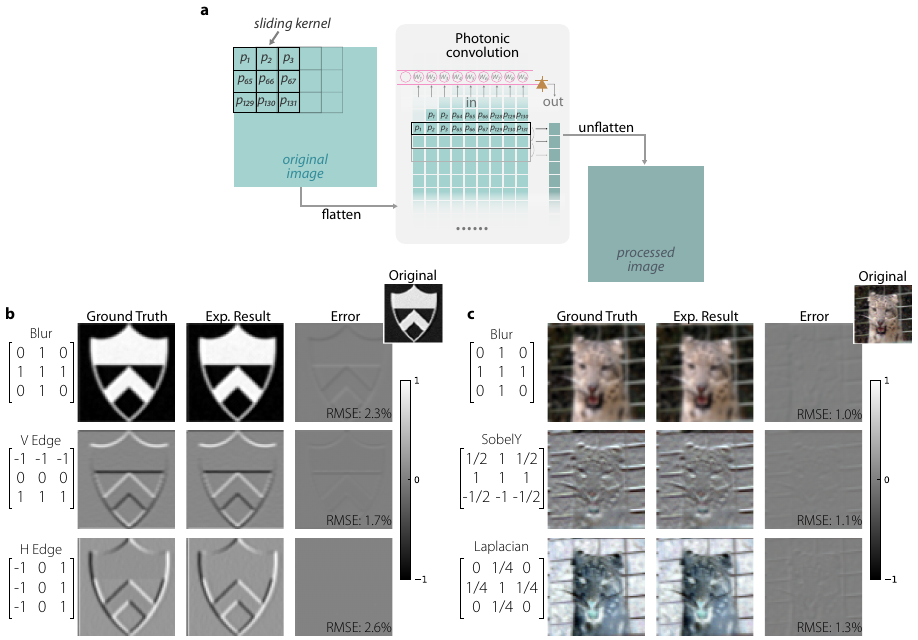}
\caption{(a) Schematic illustration of photonic image convolution. (b) Grayscale and (c) color image outputs produced by convolving each original image (upper right corner) with three different kernels, whose values are indicated in the leftmost column. In addition to the blurring effect that is achieved by averaging neighboring pixels, other kernels perform edge detection by computing local derivatives in specific directions. The Laplacian kernel applies an isotropic derivative, enhancing high-frequency features, such as the thin vertical and horizontal lines of the fence behind the tiger.
}
\label{fig:results}
\end{figure}

Image processing is a standardized benchmark task for AI hardware comparison \cite{davis2025rf, huang2022feature}. It can help validate core neural network operations such as matrix-vector multiplication \cite{pai2023experimentally, shen2017deep}, and can also highlight the parallelism and energy-efficiency of emerging physical computing platforms \cite{aadit2022massively}. To evaluate the image processing capability of the system, we configured 9 out of the 10 MRRs into a 3×3 convolutional kernel. We used a field-programmable gate array (FPGA, Xilinx RFSoC series) board with built-in DACs to generate image signals. Regarding the convolutional processing, each original image was flattened, converting it from a two-dimensional array to a one-dimensional sequence. Each MRR receives a time-shifted version of the image signal corresponding to a specific pixel within the 3×3 convolution window, implemented through staggered DAC output (Fig. \ref{fig:results}a). The time offsets are calculated based on relative pixel positions, with horizontal and vertical shifts determined by image dimensions (64×64 in this test).

Test images with a resolution of 64×64 pixels (4096 points/channel) are serialized and shaped by a raised-cosine filter (roll-off factor of 0.35), and the DACs operate at 9.8304 GS/s. The ten MRRs span a spectral window from \SI{194.2}{\tera\hertz} to \SI{195.6}{\tera\hertz}, facilitating standard wavelength-division multiplexing with ten tunable lasers combined into a single fiber for coupling into the chip. Once the system is aligned with the calibrated tuning curve of each MRR, we demonstrate classic kernels for Gaussian blurring, Sobel filtering, and the Laplacian kernel. The resulting outputs exhibit visually accurate edge maps and blurred images that closely match digital references. To quantify the fidelity, we compared the optically generated convolution output with the electronically computed ground truth using the root-mean-square error (RMSE). Fig. \ref{fig:results}b and \ref{fig:results}c present the original image, electronically computed ground truth, optical result, and corresponding pixel-wise error map. The measured RMSE ranges from 1.0\% to 2.6\%, indicating an effective MRR weight precision better than 5.3 bits and showing good accuracy for the convolution task.

Moreover, this setup exhibits very low latency, a small footprint, and modest energy consumption. The modulator and PD are integrated on the same chip, connected by a \SI{2.1}{\milli\meter} waveguide; consequently, the effective processing latency, determined by the light travel time, is on the order of tens of picoseconds. Using a single MRR for dual-function operation reduces the chip area per input channel to \SI{0.165}{\milli\meter\squared}, including bonding pads and traces. These compact and low-power characteristics confirm the viability of this photonic processing architecture. Furthermore, compatibility with WDM allows easy inclusion of additional MRRs with different resonant wavelengths. It is possible to accommodate up to 30 channels based on the current free spectral range and resonance width of each MRR. In short, our proposed architecture addresses scalability and alignment challenges in photonic processing by integrating modulation and weighting in a single MRR. The successful implementation of image processing tasks validates the practical potential of our architecture, and we anticipate that this large-scale approach will significantly expand the applicability of neuromorphic photonic processors, particularly in handling high-frequency time-series signals.

\subsection{2.4 Demonstration of High Frequency Trading}
Building on the convolutional demonstration, we now explore the capabilities of our photonic processor, realized with only 10 MRRs, for high-throughput, low-latency tasks such as real-time financial decision-making. In this section, we further enhance the proposed photonic neuron by incorporating nonlinear functionality, enabling more complex data processing with improved intelligence. Among potential real-world applications, neuromorphic photonics emerges as an ideal candidate for high-frequency trading (HFT), given its need for rapid analysis of extensive historical financial data and low-latency decision-making \cite{mackenzie2021trading}. Despite its obvious potential, to our knowledge, photonic solutions have not yet been explored for this domain.

High-frequency trading, unlike traditional portfolio management, focuses on executing a large number of trades in rapid succession, aiming to accumulate small profits that collectively yield substantial returns. A core principle in HFT is the competitive edge gained through faster and higher-quality decision-making, where even millisecond-level gains can translate into hundreds of millions of dollars for a major brokerage firm \cite{buchanan2015physics, bacidore2003order}. The primary sources of latency in HFT systems are transmission delays, dictated by geographical distance, and processing delays \cite{hasbrouck2013low}. Leading trading firms have reduced transmission time to just a few microseconds by deploying high-speed optical fiber links and relocating their infrastructure close to, or even within, the same buildings as stock exchanges \cite{menkveld2017need,zaharudin2022high,garvey2010speed}. Despite these advancements, before the emergence of photonic processors, state-of-the-art HFT systems relied on FPGAs. While FPGAs deliver the lowest latency currently achievable with electronics, they are still limited by clock speeds typically capped at a few gigahertz, resulting in latencies on the order of several microseconds \cite{subramoni2010streaming}.

Here, we demonstrate that a single photonic neuron can be configured to serve as an HFT engine with dramatically reduced latency. Neural networks are widely adopted for financial time-series prediction, capable of modeling high-dimensional, nonlinear, and noisy patterns \cite{arevalo2016high}. Compared to the mainstream linear trading algorithms implemented on FPGAs, nerual networks can learn directly from data without relying on hand-crafted rules or prior assumptions \cite{kearns2013machine}. As depicted in Fig. \ref{fig:vision}(d), the neuron can be reconfigured in a recurrent manner, using nine MRRs for input signals and one additional MRR dedicated to nonlinearity and historical feedback.

\begin{figure}[ht!]
  \centering
  \includegraphics[width=.99\linewidth]{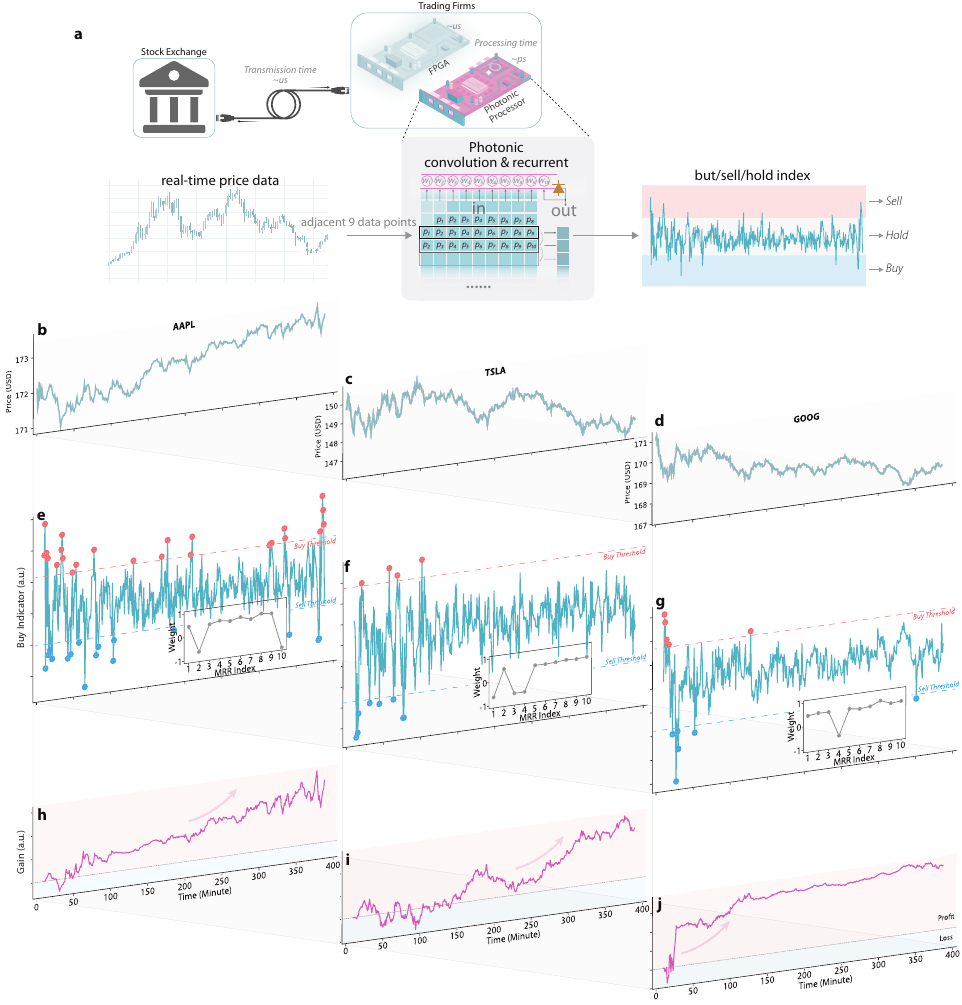}
\caption{Results of high-frequency trading. (a) Schematic illustration of the photonic high-frequency trading processor. (b–d) correspond to three different stock symbols, which are AAPL, TSLA, and GOOG, respectively. In each panel, the top row shows the stock price over one trading day. The middle row displays the simulated photonic neuron output, generated by convolving 10 consecutive price points; green and red dots denote sell and buy signals, respectively. Thresholds for buy/sell decisions are determined during PSO-based training. The bottom row illustrates the resulting cumulative profit for each stock, demonstrating robust gains irrespective of price trend.
}
\label{fig:results_hft}
\end{figure}

We evaluate the proposed system on three S\&P 500 stocks (AAPL, TSLA, and GOOG), using price data at \SI{1}{\second} intervals for the past 20 trading days, the highest temporal resolution offered by the chosen data source (Yahoo Finance). At each step, ten consecutive data points are streamed to the MRRs via the FPGA at a sampling rate of 1 GS/s. For each stock, the first 14 trading days are utilized for training, and the most recent 6 days are reserved for testing. Because the high-speed input/output interfaces of the photonic neuron are directly connected to an RFSoC FPGA that also computes trading performance via its on-board processor, we can employ online (in-situ) training to optimize the tuning currents (thereby the weights) of all ten MRRs in real-time. The training is performed using a particle swarm optimization (PSO) \cite{kennedy1995particle} algorithm, which does not require gradient calculations and converges relatively quickly in the 11-dimensional parameter space (10 MRR currents plus one bias). As discussed previously, this in-situ training effectively compensates for environmental drifts that may shift ring resonances, thus preserving optimal performance \cite{pai2023experimentally, zhang2025online}. Supplementary Fig. S8 shows experimental proof of the effectiveness of in-situ training for drift compensation, where the tuning currents were updated to recover the PNN performance under an arbitrary temperature drift from 24 \degree C to 24.5 \degree C. The training speed is limited by the electronic control loop, not the photonic hardware. Our heaters respond in 0.1 ms (10 kHz), much faster than the FPGA control latency that takes about 3 ms. A possible further improvement is to offload arithmetic to programmable logic \cite{zhang2024system}, which can approach the heater-response-dominated timescale and enable real-time adaptive updates.

Performance results of the photonic HFT system are presented in Fig. \ref{fig:results_hft}. The PD output from the recurrent network is digitized and thresholded by the FPGA to yield discrete trading decisions, namely, buy, sell, or hold, as indicated by the symbols in the middle row of Fig. \ref{fig:results_hft}. The insets in this row illustrate the final weights of the ten MRRs trained for each stock, showing how the photonic neuron converges to configurations tailored to the specific characteristics of different equities. This adaptability ultimately produces robust gains, reflected by the steadily increasing profit curves shown in the bottom row.

\begin{figure}[ht!]
  \centering
  \includegraphics[width=.99\linewidth]{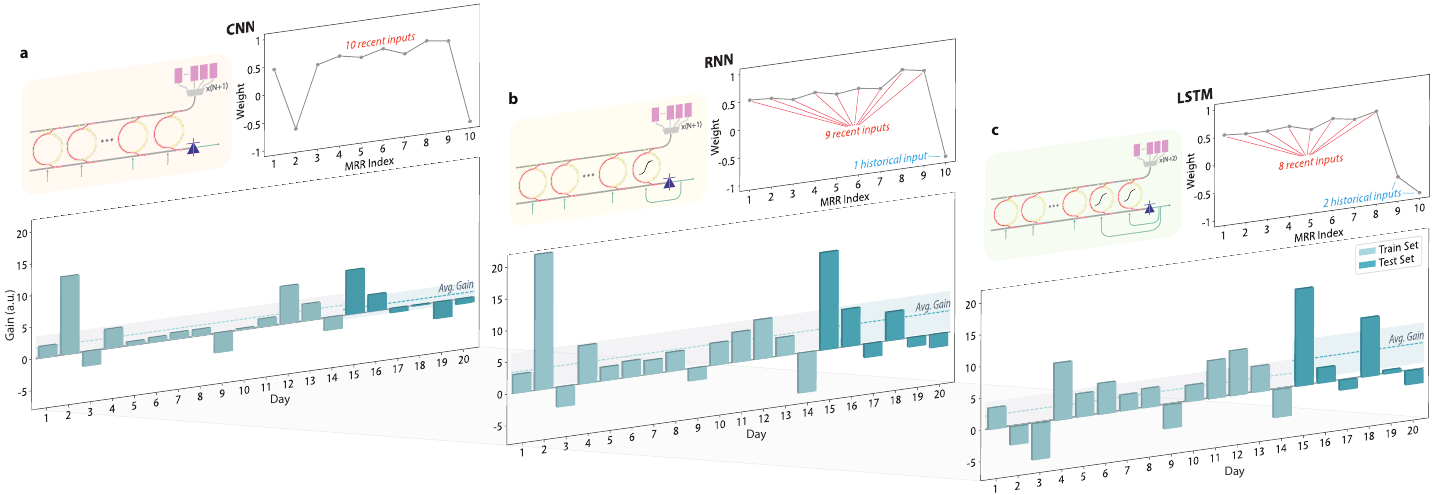}
\caption{Performance comparison of the photonic HFT using three different neuron types, which are (a) basic convolution, (b) convolution with one historical feedback path, and (c) convolution with two historical feedback paths with short and long delays.
}
\label{fig:results_hft_type}
\end{figure}

As illustrated in Fig. \ref{fig:vision}(d), the proposed modulation-and-weighting neuron accommodates a range of configurations that extend beyond convolving the 10 most recent inputs. In particular, it allows short- or long-term memory via recurrent arrangement (a feature especially beneficial for high-frequency trading tasks), where historical data may provide insights not captured by the most recent observations alone. Consequently, we evaluated our single photonic neuron under three distinct configurations for the AAPL stock symbol, as presented in Fig. \ref{fig:results_hft_type}.

The charts at the bottom depict performance over 20 consecutive trading days, with the training and testing intervals shown in light and dark colors, respectively. Although there is no guarantee of a positive return on every day, incorporating historical data clearly enhances robustness, as evidenced by reduced variability and higher gain-to-loss ratios. The charts on the right illustrate the trained weights for each configuration. For (b) and (c) that have recurrent feedback, the corresponding MRRs are trained with nominal weights around -1, which results in a sigmoid-like nonlinear transfer function, as shown in Supplementary Fig. S1(a). In all cases, the neuron effectively computes the derivative of the stock price, a strategy that, intuitively, facilitates trade-signal detection by identifying price shifts. Among the three configurations, the convolutional approach focuses on both the most recent and the oldest price derivatives, as indicated by negative weights on MRRs 2 and 10. By contrast, the other two configurations leverage historical feedback, tending to average recent data against past values, as indicated by negative weights on the final one or two MRRs. These observations underscore the versatility and emergent intelligence of the proposed photonic neuron, even when demonstrated at the relatively small scale of a single neuron.

It is worth noting that this financial time-series demonstration is a proof-of-concept rather than a direct HFT implementation, where the 1-s data rate reflects the limits of publicly accessible feeds, not of the photonic architecture. In practice, profitable HFT opportunities unfold on nanosecond timescales, which is why FPGAs currently dominate the field with few-hundred-nanosecond latencies. However, our hardware can, in principle, achieve much lower delays: the optical weight-and-sum occurs within tens of picoseconds, and all remaining latency arises from extrinsic electronics such as ADC/DAC conversion, FPGA I/O, and Python-level processing. A fully analog photonic pipeline that removes digitization, co-packages the TIA and drivers, and minimizes electrical routing could reduce end-to-end latency to sub-nanoseconds. A detailed latency breakdown and discussion can be found in Supplementary Material section 8. Further increases in modulator and PD bandwidth, together with co-designed electronics, would enable symbol rates compatible with true nanosecond-scale event processing. As neuromorphic photonic processors scale to larger networks with richer nonlinear dynamics, such architectures point toward a new computing paradigm in which market data can be processed directly as it emerges from optical fiber, bypassing conventional Ethernet stacks and electronic processors to achieve the ultralow latencies required for HFT.

\section{3 Discussions}
This work presents a large-scale, compact, and versatile photonic neuron implementation that tackles the spectral alignment challenge through a unified modulation-and-weighting MRR bank architecture. Using a single neuron composed of ten MRRs, we validate its efficacy on a classic image-processing task and demonstrate its potential in a novel real-world domain, high-frequency trading. By consolidating modulation and weighting within a single resonator, spectral alignment is substantially simplified, enabling a straightforward 10-MRR configuration. The versatility is further enhanced by its configurability; beyond the standard feedforward approach, any subset of MRRs can handle linear weighted summation, while the remainder provides nonlinear activation with adjustable delays. This flexibility enables the same MRR bank to support recurrent operation via an electrical feedback path, thereby enabling memory functions that allow the neuron to process temporal signals, such as financial time-series data.

From a quantitative perspective, each modulation-weighting element, including the microring, heater, and basic electrode, occupies only \SI{80}{\micro\meter} by \SI{45}{\micro\meter}. Operating at 9.83 GS/s for all 10 MRRs, the total compute speed is rated as \SI{0.196}{TOPs/s}. Given the footprint of the PD to be \SI{120}{\micro\meter} by \SI{50}{\micro\meter}, the total required chip area for this setup is \SI{0.042}{\milli\meter^2}, and therefore the compute density is \SI{4.67}{TOPs/s/\milli\meter^2}. As for power consumption, the total power required by on-chip tuning is dominated by the thermal-optic tuning of the MRRs, which amounts to \SI{0.186}{\milli\watt} on average. Thus, our neuron achieves an energy efficiency of \SI{105}{TOPs/\watt}. A detailed power breakdown analysis is summarized in Supplementary Table S1. To contextualize the performance of our architecture, Supplementary Table S2 provides a quantitative comparison with representative state-of-the-art integrated photonic neural and neuromorphic processors \cite{shen2017deep, xu202111, feldmann2021parallel, pai2023experimentally, bai2023microcomb, meng2023compact, zheng2024photonic, xie2025complex, he2025programmable}.

Regarding the scalability to even larger arrays, it primarily requires addressing fabrication variation via laser tuning or trimming, enhancing thermal isolation through deep-trench structures or optimized heaters, and managing electrical routing using traveling-wave electrodes and advanced packaging. Because the architecture operates at low optical power and sub-mW heater levels, the laser power budget and thermal load remain modest, and established photonic-integration techniques provide a clear path toward arrays with tens of weights. With these, we anticipate that this newly enhanced photonic neuron architecture will serve as a scalable building block for future neuromorphic photonic systems, combining compact footprint, high compute density, ultra-low power consumption, and reconfigurability to address the demands of large-scale neuromorphic computing in real-world applications.

\section{4 Methods}

\subsection{4.1 Photonic chip packaging}
The PIC, along with two TIAs and several decoupling capacitors, is mounted directly onto the top PCB using silver epoxy and wire-bonded to the PCB pads for electrical interconnects. Rogers 4003C serves as the PCB substrate to accommodate high-frequency signals with minimal loss. RF signals enter and exit the PCB via SMP connectors and traverse impedance-controlled traces to the PIC. Optical coupling is achieved through a glued fiber array, polished at a \SI{41}{\degree} angle to match the on-chip grating couplers, which are optimized for an 8-degree incident angle, yielding a measured coupling loss of approximately \SI{7}{\decibel} per fiber-to-chip pass.

A ribbon cable connects the top PCB to the bottom PCB, which supplies tuning currents to the MRR banks and bias voltages to the MRRs and PDs. This bottom PCB is populated with two five-channel current-output DACs (LTC2662, Analog Devices) and one 8-channel voltage-output DAC (LTC2686, Analog Devices), along with several low-noise linear power modules to generate the required \SI{+5}{\volt}, \SI{+10}{\volt}, and \SI{-10}{\volt} rails. The bottom PCB also interfaces with the FPGA board via a high-speed Serial Peripheral Interface (SPI) protocol for configuration and control. Additionally, a thermoelectric cooler is located between the two PCBs, working in conjunction with a thermistor on the top board and a dedicated temperature controller to maintain thermal stability.

\subsection{4.2 Compact optical coupling}
Optical coupling requires precise alignment between the external fiber array and the on-chip grating couplers. This process involves a trade-off between permanent solutions (e.g., epoxy bonding, photonic wire bonding), which lack replaceability, and bulky temporary setups (e.g., multi-axis alignment stages), which can be cumbersome. To address this challenge, we developed a compact alignment mechanism for temporary optical packaging that is compatible with grating couplers. Our design employs a Thorlabs optical cage system consisting of two primary mounting plates connected by four rods. The interposer PCB carrying the photonic chip is affixed to the bottom plate (Thorlabs CXYZ05A), enabling rotation about the z-axis. The fiber array is secured to the top plate (Thorlabs CRM1PT) via a custom adapter, which provides translational motion in the x, y, and z directions. The adapter features an inclined mounting plane set at 37° to the x-y plane, matching the angle required by the grating couplers.

Using an on-chip loop waveguide that routes one grating coupler directly to another, we measured a round-trip loss of approximately 11 dB and observed a long-term drift of less than 0.05 dB. Accounting for the inherent waveguide losses, this corresponds to a coupling loss better than 5.5 dB, being comparable to bulkier setups with more sophisticated alignment mechanisms \cite{zhang2022silicon}.

\subsection{4.3 FPGA-based Signal Generation}
The FPGA board used in this setup is a Xilinx RFSoC-series model that provides a multi-channel, high-speed, and cost-effective solution for both signal generation (digital-to-analog conversion) and signal digitization (analog-to-digital conversion). Specifically, we employed the HTG-ZRF16 board from HiTech Global, featuring a ZU49DR RFSoC FPGA chip. This board can be programmed using either Xilinx’s official Vivado software or the third-party CasperFPGA firmware, developed by the Collaboration for Astronomy Signal Processing and Electronics Research (CASPER), a consortium of multiple research groups. The FPGA program utilized in this work was derived from CasperFPGA, and we acknowledge the CASPER organization for making their code openly available.

\section{Acknowledgments}
This research is supported by the National Science Foundation (NSF) (ECCS-2128616 and ECCS-1642962 to P.R.P.), the Office of Naval Research (ONR) (N00014-18-1-2297 and N00014-20-1-2664 to P.R.P.), and the Defense Advanced Research Projects Agency (HR00111990049 to P.R.P.). The devices were fabricated at the Advanced Micro Foundry (AMF) in Singapore through the support of CMC Microsystems. B. J. Shastri acknowledges support from the Natural Sciences and Engineering Research Council of Canada (NSERC).

\section{Competing interests}
The authors declare that they have no competing interests

\section{Availability of data and materials}
The datasets used and/or analysed during the current study are available from the corresponding author on reasonable request.

\section{Author contributions}
W.Z. conceived the ideas. W.Z. and Y.W. designed the experiment and conducted the experimental measurements. W.Z. analyzed the results. J.C.L. designed the silicon photonic chip. B.J.S. provided theoretical support. W.Z., Y.W. and B.J.S. wrote the manuscript. P.R.P. supervised the research and contributed to the general concept and interpretation of the results. All the authors discussed the data and contributed to the manuscript.

\backmatter


\begin{thebibliography}{48}

\bibitem{brunner2025roadmap} D. Brunner et al., Roadmap on neuromorphic photonics. Preprint at https://arxiv.org/abs/2501.07917 (2025).

\bibitem{zhang2024system} W. Zhang et al., A system-on-chip microwave photonic processor solves dynamic RF interference in real time with picosecond latency. Light Sci. Appl. \textbf{13}, 14 (2024).

\bibitem{shekhar2024roadmapping} S. Shekhar et al., Roadmapping the next generation of silicon photonics. Nat. Commun. \textbf{15}, 751 (2024).

\bibitem{huang2022prospects} C. Huang et al., Prospects and applications of photonic neural networks. Adv. Phys. X \textbf{7}, 1981155 (2022).

\bibitem{fu2024optical} T. Fu et al., Optical neural networks: progress and challenges. Light Sci. Appl. \textbf{13}, 263 (2024).

\bibitem{xu2023integrated} X. Xu and X. Jin, Integrated photonic computing beyond the von neumann architecture. ACS Photonics \textbf{10}, 1027--1036 (2023).

\bibitem{ferreira2017progress} T. Ferreira De Lima, B.J. Shastri, A.N. Tait, M.A. Nahmias, and P.R. Prucnal, Progress in neuromorphic photonics. Nanophotonics \textbf{6}, 577--599 (2017).

\bibitem{xu2024control} T. Xu et al., Control-free and efficient integrated photonic neural networks via hardware-aware training and pruning. Optica \textbf{11}, 1039--1049 (2024).

\bibitem{ferreira2022design} E.A. Doris et al., Design automation of photonic resonator weights. Nanophotonics \textbf{11}, 3805--3822 (2022).

\bibitem{bogaerts2018silicon} W. Bogaerts and L. Chrostowski, Silicon photonics circuit design: methods, tools and challenges. Laser Photon. Rev. \textbf{12}, 1700237 (2018).

\bibitem{tait2017neuromorphic} A.N. Tait et al., Neuromorphic photonic networks using silicon photonic weight banks. Sci. Rep. \textbf{7}, 7430 (2017).

\bibitem{bogaerts2013design} W. Bogaerts, M. Fiers, and P. Dumon, Design challenges in silicon photonics. IEEE JSTQE \textbf{20}, 1--8 (2013).

\bibitem{zhang2022silicon} W. Zhang et al., Silicon microring synapses enable photonic deep learning beyond 9-bit precision. Optica \textbf{9}, 579--584 (2022).

\bibitem{tait2016microring} A.N. Tait et al., Microring weight banks. IEEE JSTQE \textbf{22}, 312--325 (2016).

\bibitem{bogaerts2012silicon} W. Bogaerts et al., Silicon microring resonators. Laser Photon. Rev. \textbf{6}, 47--73 (2012).

\bibitem{vaswani2017attention} A. Vaswani, Attention is all you need. Paper presented at 31st Annual Conference on Neural Information Processing Systems (NIPS), Long Beach, 4-9 Dec 2017.

\bibitem{kaplan2020scaling} J. Kaplan et al., Scaling laws for neural language models. Preprint at https://arxiv.org/abs/2001.08361 (2020).

\bibitem{huang2020demonstration} C. Huang et al., Demonstration of scalable microring weight bank control for large-scale photonic integrated circuits. APL Photonics \textbf{5}, 4 (2020).

\bibitem{zhang2024microring} W. Zhang, J. Zhang, J.C. Lederman, B.J. Shastri, and P. Prucnal, Microring modulation-and-weight banks. Paper presented at 44th Conference on Lasers and Electro-Optics (CLEO), Charlotte, 5-10 May 2024.

\bibitem{li2021survey} Z. Li, F. Liu, W. Yang, S. Peng, and J. Zhou, A survey of convolutional neural networks: analysis, applications, and prospects. IEEE Trans. Neural Netw. Learn. Syst. \textbf{33}, 6999--7019 (2021).

\bibitem{xu202111} X. Xu et al., 11 tops photonic convolutional accelerator for optical neural networks. Nature \textbf{589}, 44--51 (2021).

\bibitem{tait2019silicon} A.N. Tait et al., Silicon photonic modulator neuron. Phys. Rev. Appl. \textbf{11}, 064043 (2019).

\bibitem{tait2022quantifying} A.N. Tait, Quantifying power in silicon photonic neural networks. Phys. Rev. Appl. \textbf{17}, 054029 (2022).

\bibitem{narayana2017morphonoc} V.K. Narayana, S. Sun, A.-H.A. Badawy, V.J. Sorger, and T. El-Ghazawi, Morphonoc: Exploring the design space of a configurable hybrid noc using nanophotonics. Microprocess. Microsyst. \textbf{50}, 113--126 (2017).

\bibitem{dong2023higher} B. Dong et al., Higher-dimensional processing using a photonic tensor core with continuous-time data. Nat. Photon. \textbf{17}, 1080--1088 (2023).

\bibitem{lederman2023real} J.C. Lederman et al., Real-time photonic blind interference cancellation. Nat. Commun. \textbf{14}, 8197 (2023).

\bibitem{zhang2023broadband} W. Zhang et al., Broadband physical layer cognitive radio with an integrated photonic processor for blind source separation. Nat. Commun. \textbf{14}, 1107 (2023).

\bibitem{davis2025rf} R. Davis III, Z. Chen, R. Hamerly, and D. Englund, RF-photonic deep learning processor with Shannon-limited data movement. Sci. Adv. \textbf{11}, 3558 (2025).

\bibitem{huang2022feature} Y. Huang et al., Feature extraction from images using integrated photonic convolutional kernel. IEEE Photon. J. \textbf{14}, 1--7 (2022).

\bibitem{pai2023experimentally} S. Pai et al., Experimentally realized in situ backpropagation for deep learning in photonic neural networks. Science \textbf{380}, 398--404 (2023).

\bibitem{shen2017deep} Y. Shen et al., Deep learning with coherent nanophotonic circuits. Nat. Photon. \textbf{11}, 441--446 (2017).

\bibitem{aadit2022massively} N.A. Aadit et al., Massively parallel probabilistic computing with sparse ising machines. Nat. Electron. \textbf{5}, 460--468 (2022).

\bibitem{mackenzie2021trading} D. MacKenzie, \textit{Trading at the speed of light: How ultrafast algorithms are transforming financial markets} (Princeton University Press, Princeton, 2021), pp. 1-304.

\bibitem{buchanan2015physics} M. Buchanan, Physics in finance: Trading at the speed of light. Nature \textbf{518}, 161--163 (2015).

\bibitem{bacidore2003order} J. Bacidore, R.H. Battalio, and R.H. Jennings, Order submission strategies, liquidity supply, and trading in pennies on the new york stock exchange. J. Financ. Mark. \textbf{6}, 337--362 (2003).

\bibitem{hasbrouck2013low} J. Hasbrouck and G. Saar, Low-latency trading. J. Financ. Mark. \textbf{16}, 646--679 (2013).

\bibitem{menkveld2017need} A.J. Menkveld and M.A. Zoican, Need for speed? exchange latency and liquidity. Rev. Financ. Stud. \textbf{30}, 1188--1228 (2017).

\bibitem{zaharudin2022high} K.Z. Zaharudin, M.R. Young, and W.-H. Hsu, High-frequency trading: Definition, implications, and controversies. J. Econ. Surv. \textbf{36}, 75--107 (2022).

\bibitem{garvey2010speed} R. Garvey and F. Wu, Speed, distance, and electronic trading: New evidence on why location matters. J. Financ. Mark. \textbf{13}, 367--396 (2010).

\bibitem{subramoni2010streaming} H. Subramoni, F. Petrini, V. Agarwal, and D. Pasetto, Streaming, low-latency communication in on-line trading systems. Paper presented at 24th IEEE International Symposium on Parallel \& Distributed Processing, Workshops and Phd Forum (IPDPSW), Georgia, 19-23 Apr 2010.

\bibitem{arevalo2016high} J. Sandoval, High-frequency trading strategy based on deep neural networks, Paper presented at 12th International conference on intelligent computing (ICIC), Lanzhou, 2-5 Aug 2016.

\bibitem{kearns2013machine} M. Kearns and Y. Nevmyvaka, in Machine learning for market microstructure and high frequency trading, ed. by M. O'hara, M. López de Prado and D. Easley. \textit{High Frequency Trading: New Realities for Traders, Markets, and Regulators}, vol 72 (Risk Books, London, 2013) p. 1877.

\bibitem{kennedy1995particle} J. Kennedy and R. Eberhart, Particle swarm optimization. Paper presented at the International conference on neural networks (ICNN), Perth, 27-31 Nov 1995.

\bibitem{zhang2025online} J. Zhang et al., Online training and pruning of multi-wavelength photonic neural networks. Nanophotonics \textbf{14}, 5035--5046 (2025).

\bibitem{feldmann2021parallel} J. Feldmann et al., Parallel convolutional processing using an integrated photonic tensor core. Nature \textbf{589}, 52--58 (2021).

\bibitem{bai2023microcomb} B. Bai et al., Microcomb-based integrated photonic processing unit. Nat. Commun. \textbf{14}, 66 (2023).

\bibitem{meng2023compact} X. Meng et al., Compact optical convolution processing unit based on multimode interference. Nat. Commun. \textbf{14}, 3000 (2023).

\bibitem{zheng2024photonic} Y. Zheng et al., Photonic neural network fabricated on thin film lithium niobate for high-fidelity and power-efficient matrix computation. Laser Photon. Rev. \textbf{18}, 2400565 (2024).

\bibitem{xie2025complex} Y. Xie et al., Complex-valued matrix-vector multiplication using a scalable coherent photonic processor. Sci. Adv. \textbf{11}, 7475 (2025).

\bibitem{he2025programmable} J. He et al., Programmable electro-optic frequency comb empowers integrated parallel convolution processing. Preprint at https://arxiv.org/abs/2506.18310 (2025).

\end{thebibliography}
\end{document}